\documentclass[aps,amsmath,amssymb,12pt]{revtex4}
\usepackage{graphicx}
\usepackage{bm}
\usepackage{tcolorbox}

\def\journal#1#2#3#4{{\em #1} {\bf #4}, \emph{#2}, #3}
\newcommand{\be}{\begin{equation}}
\newcommand{\ee}{\end{equation}}
\newcommand{\bea}{\begin{eqnarray}}
\newcommand{\eea}{\end{eqnarray}}
\newcommand{\hf}{\frac12}
\newcommand{\nn}{\cr}
\def\eq#1{(\ref{#1})}
\def\la{\langle}
\def\ra{\rangle}

\def\Tr{{\mathrm{Tr}}}
\def\ord#1{{\cal O}\left(#1\right)}
\def\mr#1{{\mathrm{#1}}}
\def\v#1{{\bm{#1}}}

\def\fdd#1#2#3{\frac{\delta^2#1}{\delta#2\delta#3}}
\def\dk{{\Delta k}}
\def\hphi{\hat\phi}
\def\hj{\hat j}

\def\hD{{\hat D}}

\def\sign{\mr{sign}}

\begin{document}
\title{Renormalizing open quantum field theories}

\author{S. Nagy$^1$, J. Polonyi$^2$}
\affiliation{$^1$Department of Theoretical Physics, University of Debrecen, P.O. Box 5, H-4010 Debrecen, Hungary}
\affiliation{$^2$Strasbourg University, CNRS-IPHC, \\23 rue du Loess, BP28 67037 Strasbourg Cedex 2, France}
\date{\today}

\begin{abstract}
The {functional} renormalization group flow of a scalar field theory with quartic couplings and a sharp {spatial} momentum cutoff is presented in four-dimensional Minkowski space-time {for the bare action} by retaining the entanglement of the IR and the UV particle modes. It is argued that the open interaction channels have to be taken into account in quantum field theory defined by the help of a cutoff, and a non-perturbative UV-IR entanglement is found in closed or almost closed models.
\end{abstract}

\maketitle

\section{Introduction}
One of the most important lessons of contemporary physics is that the observed phenomena depend qualitatively and quantitatively on the scales of observation. Hence rather than looking for a ``World Equation'' or ``Theory Of Everything'' one looks for effective theories, valid only in some scale window. We know several such theories, starting with the Standard Model of Particle Physics and ending with the Standard Model of Cosmology, and the challenge is to understand the way these models follow each other on their renormalized trajectory as the resolution of the observations is changed. The~goal of this work is to orient the attention to an important feature of the effective theories, namely that they deal with open dynamics. However, this point is actually obvious, since the unobserved degrees of freedom represent an environment for the observed system the systematic addressing of the problem has been lacking. {We choose the simplest non-trivial model for this purpose, describing a scalar field in 3 + 1 dimensions.}

There is another reason to consider open theories. In~the first phase of the history of quantum field theory,  attention was turned towards renormalizable models by the help of renormalized perturbation expansion. However, the need to go beyond this approximation scheme introduced the cutoff theories, which are defined by a large but finite UV cutoff back into the foreground of the interest. The~cutoff theories describe open dynamics, as~well, since the degrees of freedom beyond the cutoff serve as an environment. One can go a bit further an state that the inherent UV divergences of quantum field theory simply exclude a truly closed quantum dynamics by rendering the cutoff necessary. One might object that an environment consisting of very energetic particle modes should not modify the low energy physics in an important manner. However the question is more involved and a detailed knowledge of the scaling laws are needed to understand the role of the open channels at high energy in the physics at low~energy.

Our main results, obtained for the four-dimensional real scalar theory, are as follows: (i) The change of the cutoff towards either the IR or the UV direction renders the dynamics open, in~other words, closed theories are inconsistent according to the renormalization group method. {This is demonstrated explicitly in Section~\ref{btrs}.} (ii) There are open interaction vertices representing open interactions which are relevant and leave a trace on the physical quantities for arbitrarily high cutoff scale. (iii) The theory with sufficiently long lived quasi particles displays non-perturbative IR scaling and strong UV-IR entanglement, making the comparison of the quantum and the classical dynamics more difficult. Therefore quantum field theories should be used by allowing a mixed vacuum state. The~most promising method to deal with open systems is the Closed Time Path (CTP) scheme (or Schwinger-Keldysh formalism) hence the renormalization of effective quantum field theories should be handled in that~formalism. 

The CTP formalism was first  developed for the perturbation expansion in the Heisenberg representation~\cite{schwe,schwk,keldysh}; however its subsequent use is increasingly in open quantum systems where the system-environment separation appears in different disguise in different physical problem. The~decoherence~\cite{zehd,zurekd}, a~necessary condition of the classical limit of quantum systems~\cite{joos,zurekt}, can easily be grasped by establishing the additive probabilities of histories~\cite{gellmann,griffiths,omnes,halliwell} in a macroscopic environment. The~environment of an observed collective mode consists of the rest of the macroscopic system in non-equilibrium statistical physics~\cite{kamenev,rammer}. The~environment of a dissipative system remains unreachable~\cite{weiss,zaikin}. The~environment of driven nanophysical, solid state or optical devices is in the macroscopic domain~\cite{sieberer}. The~nano wires are imbedded into an environment of their leads~\cite{bertini}. A~thermal reservoir can always be assumed as part of the environment, too. High energy physics applications stretch from thermal field theory~\cite{umezawa} to astrophysics~\cite{calzetta}. The~path integral representation of quantum mechanics is particularly well suited to this formalism~\cite{feynman}, the~calculation we report below has been performed in the framework of the path integral representation of the CTP formalism, covered concisely in ref.~\cite{calzetta}. The~environment of a cutoff field theory is rather particular, it is provided by the UV modes which interact with the retained IR modes. Our goal is to discover the impact of the entanglement between the IR and the UV sectors on the IR~dynamics.

The functional renormalization group method was already applied in the CTP formalism to follow the coarse graining~\cite{lombardo,dalvit,anastopoulos}, addressing a quantum dot~\cite{gezzi}, open electronic systems~\cite{mitra}, the~transport processes~\cite{jacobs}, the~damping~\cite{zanellad}, the~inflation~\cite{zanellai}, quantum cosmology~\cite{calzettac} and critical dynamics~\cite{bergerhoff,canet,mesterhazy,sieberere,siebererk}. Furthermore, it can describe the behavior of the Bose--Einstein condensate~\cite{gasenzerbe,berges,siebererh}, the~form of the spectral function~\cite{pawlowskisf,huelsmann}, or~real time dynamics of gauge theories~\cite{kasper}. The~renormalization group scheme was extended to stochastic field theory~\cite{zanellac}, too. The~need for bi-local terms in the action was argued in ref.~\cite{qrg}. The~extension to the 2PI formalism has been used used to find non-thermal fixed points~\cite{bergesfp} and the renormalization group scheme can be transformed to trace the time dependence~\cite{gasenzer,gasenzerk,corell}. The~one-loop renormalizability of the scalar model has been worked out on the one-loop level by the help of the more traditional multiplicative renormalization group method~\cite{avinash,avinashk}.

Before starting, it is worth distinguishing between a regulator and a cutoff of a field theory. The~former renders the theory UV finite and the latter introduces a separation of the UV and the IR modes. A~regulated theory, which is formulated in continuous space-time at arbitrarily large frequencies and wave vectors, can be called microscopic and a theory with a cutoff is necessarily effective. A~cutoff of scale $\Lambda$ leaves the {field modes with spatial three-momentum $|\v{p}|<(1-\delta_-)\Lambda$ untouched in the IR} and suppresses them completely in the UV for $|\v{p}|>(1+\delta_+)\Lambda$ where $0\le\delta_-<1$ and $0\le\delta_+<\infty$. The~modes $(1-\delta_-)\Lambda<|\v{p}|<(1+\delta_+)\Lambda$ are neither IR nor UV. A~cutoff with $\delta_\pm=0$ is called sharp and yields a well defined splitting of the degrees of freedoms into an IR and an UV class. A~smooth cutoff with $\delta_+=\infty$ is only a regulator. In~a regulated (UV finite) theory without cutoff the separation of the UV and the IR sector can be defined only qualitatively by introducing the UV regime for scales where the dynamics is strongly influenced by the~regulator.

The choice of the width of a smooth cutoff, $\delta_-+\delta_+$, represents a compromise. One the one hand, the~width should be small to separate clearly the observed and the unobserved modes. On~the other hand, the~width should be large enough to avoid long range oscillations in the space-time. According to the traditional strategy, one uses the truncated gradient expansion as an ansatz for the action and such oscillations make  this approximation scheme ill-defined even in Euclidean space-time where the dynamics is free of mass-shell singularities. However, the gradient expansion is a dead end street in constructing a systematic approximation to real time dynamics owing to Ostrogadsky's instability~\cite{ostrogadsky,woodard} and should be replaced by the cluster expansion based on multi-local action functionals. By~anticipating such a strategy for the future, we rely on the functional renormalization group method based on sharp cutoff { with $\delta_\pm=0$} \cite{wh}, used in the CTP formalism in the present~work.

The CTP scheme is introduced briefly in Section~\ref{oqss} below and the implementation of the gliding cutoff follows in Section~\ref{btrs}. The~renormalization group trajectory is discussed in Section~\ref{rgflows}. The~summary of our result together with the conclusion are presented in Section~\ref{concl}.

\section{Open Quantum~Systems}\label{oqss}
The lowering of the cutoff, the~blocking, has to be performed by keeping track of the mixed state components generated by the elimination of the dynamical degrees of freedom. This can be achieved in the CTP formalism, outlined in this~section.

\subsection{Closed~System}
We start with the density matrix of a closed system,
\be\label{densm}
\rho[t_f,\Phi_+,\Phi_-]=\la\Phi_+|U(t_f,t_i)\rho(t_i)U^\dagger(t_f,t_i)|\Phi_-\ra,
\ee
where $\rho(t_i)$ stands for the initial density matrix and $U(t_f,t_i)$ denotes the time evolution operator of a closed dynamics. The~path integral expression,
\be\label{genfpi}
\rho=\int D[\hphi]e^{iS[\hphi]},
\ee
can be obtained by performing the usual slicing procedure in time for $U$ and $U^\dagger$ where $\hphi=(\phi_+,\phi_-)$ denotes the doublet field, the~integration is over the field configurations $\phi_\pm(t_f,\v{x})=\Phi_\pm(\v{x})$, the~convolution with the initial density matrix at time $t_i$ is suppressed in the notation for simplicity and the action is given by $S[\hphi]=S[\phi_+]-S^*[\phi_-]$, $S[\phi]$ denoting the usual action of the closed~theory. 

The traditional formalism of quantum field theory dealing with transition amplitudes between pure states, $\la\Phi_f|U(t_f,t_i)|\Phi_i\ra$, is called Single Time Path (STP) scheme. The~reduplication of the degrees of freedom in \eq{genfpi} is to cope with the quantum fluctuations of the bra and the ket in \eq{densm}. This is not necessary in closed dynamics for pure initial states, $\la\Phi_+|\rho(t_i)|\Phi_-\ra=\Psi_i(\Phi_+)\Psi_i^*(\Phi_-)$, where these fluctuations are independent and identical. The~expression \eq{genfpi} of the density matrix belongs to the Open Time Path (OTP) scheme because the trajectories of the path integral are open, have different end points. The~reduplication of the degrees of freedom is unavoidable in open systems where the non-factorizable density matrix of a mixed state describes correlated bra and ket~fluctuations.

\subsection{Open~System}
Let us now assume that we observe the dynamics of the field $\phi$ which is interacting with another field, $\varphi$, the~full dynamics is closed and is described by the action $S[\phi,\varphi]=S_s[\phi]+S_e[\phi,\varphi]$. To assure the reader that the environment influences the observed system by the interactions taking place only during the observation time we assume that the system and its environment are not entangled at the initial time meaning that the initial density matrix factorizes as $\rho(t_i)=\rho_{s,0}(t_i)\rho_{e,0}(t_i)$. We are interested in the reduced density matrix of the observed system,
\be\label{densm}
\rho[t_f,\Phi_+,\Phi_-]=\la\Phi_+|\Tr_\varphi[U(t_f,t_i)\rho(t_i)U^\dagger(t_f,t_i)]|\Phi_-\ra,
\ee
given by the path integral expression, \eq{genfpi}, and~the effective action, $S[\hphi]=S_s[\phi_+]-S_s^*[\phi_-]+S_{infl}[\hphi]$. The~influence functional describes the open interactions~\cite{feynman},
\be\label{ctpint}
e^{iS_{infl}[\hphi]}=\int D[\hat\varphi]e^{iS_e[\phi_+,\varphi_+]-iS^*_e[\phi_-,\varphi_-]},
\ee
where the integration is over the field configurations $\varphi_+(t_f,\v{x})=\varphi_-(t_f,\v{x})$ and the convolution with the initial density matrix is suppressed represents the system-environment~interactions.

It is advantageous to write the resulting effective action in the form 
\be\label{effctpa}
S[\hphi]=S_1[\phi_+]-S_1^*[\phi_-]+S_2[\phi_+,\phi_-],
\ee
by separating the uncoupled and the coupled time axis contributions, $\delta^2S_2/\delta\phi_+\delta\phi_-\ne0$.  The~independence of the bra and the ket fluctuations is reflected in the simple additive structure of the action and the single time axes contribution, $S_1$, comprises the closed, conservative interactions. The~coupling between the axis, $S_2$, generates open classical forces, correlates the bra and ket fluctuations and renders the reduced density matrix mixed~\cite{effthe}.

The physical content of the effective theory can be extracted from the generator functional for the connected Green functions, 
\be\label{effgenfunct}
e^{iW[\hj]}=\Tr_\phi\Tr_\varphi[U(t_f,t_i;j_+)\rho(t_i)U^\dagger(t_f,t_i;-j_-)]
\ee
where the system dynamics is extended by the introduction of the external source $j(x)$, coupled linearly to the field $\phi(x)$ in the action, giving rise to the time evolution operator $U(t_f,t_i;j)$. This equation defines the CTP scheme because the path integral expression of the generator functional contains closed trajectory pairs owing to the trace over the Fock space. The~shifts $t_i\to t_i+\tau$ and $t_f\to t_f+\tau$ is a symmetry of the dynamics due to the time translation invariance of the action. This symmetry, which is important in the STP formalism to keep the Green functions diagonal in the frequency space, is violated by the trace operation. To~regain the symmetry and simplify the formalism, we perform the long evolution time limit, $t_i\to-\infty$, $t_f\to\infty$, which renders the propagator of local excitations diagonal in the continuous frequency space. However this step is more involved as in the STP case~\cite{irr}. In~fact, this limit is assured in the latter case by Feynman's $i\epsilon$ prescription, the~adiabatic suppression of the excitations for long time evolution. Rather than approaching slowly the vacuum, the~final state is allowed to be chosen freely by the dynamics itself in the CTP formalism hence the dynamics is non-trivial at any finite $t_f$. 

Finally, a comment on the time reversal invariance. The~limit $t_i\to-\infty$ and $t_f\to\infty$ simplifies the time reversal transformation to $S[\phi_+,\phi_-]\to T(S[\phi_+,\phi_-])=-S^*[\phi_-,\phi_+]$, namely the CTP generator functional with an action $S[\phi_+,\phi_-]=-S^*[\phi_-,\phi_+]$ and without  any particular rule at the initial and the final time is time reversal invariant. The~action of a closed dynamics is time reversal invariant in this limit. The~influence functional \eq{ctpint} preserves this symmetry therefore open dynamics remains time reversal, as~well. There is no contradiction with the presence of dissipative forces which may arise in an open dynamics since the time reversal, as~defined above, reverses the direction of the time both for the system and for its environment. In~particular, the~coupling between the time axis generates terms in the (effective) equation of motion with broken time reversal symmetry~\cite{effthe}.

\subsection{Propagator}
The limits $t_i\to-\infty$ and $t_f\to\infty$ can easily be found for free quantum fields in the following manner. The~dynamics defined by the translation invariant action
\be\label{freectpa}
S_0=\hf\int dxdy\hphi(x)\hD^{-1}(x-y)\hphi(y)
\ee
yields the generator functional
\be
e^{iW[\hj]}=\int D[\hphi]e^{iS_0[\hphi]+i\int dx\hj(x)\hphi(x)}=e^{-\frac{i}2\int dxdy\hj(x)\hD(x-y)\hj(y)}
\ee
with the CTP propagator,
\be
\fdd{iW[\hj]}{i\hj(x)}{i\hj(y)}=i\hD(x-y)
=\begin{pmatrix}\Tr[T[\phi_+(x)\phi_+(y)]\rho]&\Tr[\phi_-(y)\phi_+(x)\rho]\cr\Tr[\phi_-(x)\phi_+(y)\rho]&\Tr[T[\phi_-(y)\phi_-(x)]^*\rho]\end{pmatrix},
\ee
containing the Feynman propagator and its complex conjugate in the diagonal and the Wightman function in its off-diagonal blocks. First one calculates these functions in the operator formalism in the limit $t_i\to-\infty$ and $t_f\to\infty$ for the vacuum as initial state, 
\be\label{frprop}
\hD_p=\int dxe^{ip(x-y)}\hD(x-y)
=\begin{pmatrix}\frac1{p^2-m^2+i\epsilon}&-2i\pi\delta(p^2-m^2)\Theta_{-p^0}\cr-2i\pi\delta(p^2-m^2)\Theta_{p^0}&-\frac1{p^2-m^2-i\epsilon}\end{pmatrix}
\ee
for a scalar particle of mass $m$. Please note that the only source of the off-shell amplitudes in the propagator is the time ordering, the~Wightman functions are on-shell and the Heaviside functions represent the vacuum initial condition. Next the kernel of the free action, the~inverse propagator is found,
\be\label{invfrpr}
\hD_p^{-1}=\begin{pmatrix}p^2-m^2+i\epsilon&-2i\epsilon\Theta_{-p^0}\cr-2i\epsilon\Theta_{p^0}&m^2-p^2+i\epsilon
\end{pmatrix},
\ee
by direct inversion of \eq{frprop} when the Dirac-delta distribution is represented by a regulated normalized Lorentzian~peak.

The doublet fields $\phi_\pm(x)$ are coupled only at the final time for finite $t_f$. One would have thought that this coupling becomes unimportant in the limit $t_f\to\infty$. However, such a naive argument is misleading, since the suppression of the coupling between the doublets removes the trace in \eq{effgenfunct} for whatever large $t_f$ we use. According to \eq{invfrpr}, the limit $t_f\to\infty$ indeed suppresses the coupling at $t_f$ and renders the dynamics translation invariant in time, but the two time axes remain coupled by an infinitesimal, time translation invariant operator, the~off-diagonal elements on the right had side of \eq{invfrpr}. This coupling represents the difference between the CTP and the STP schemes and is non-local in time,
\bea
\tilde\Theta(t-t')&=&\int\frac{d\omega}{2\pi}e^{-i\omega(t-t')}\Theta_\omega\nn
&=&-\frac{i}{t-t'-i\epsilon},
\eea
to assure that the elementary excitations above the vacuum, propagating with \eq{frprop}, correspond to positive~energies.

\subsection{Initial~State}\label{insts}
The initial state $\rho(t_i)$ in \eq{densm} influences the effective action for the observed subsystem. We choose the perturbative vacuum as initial state in the limit $t_i\to-\infty$ and assume the usual adiabatic building up the true, interactig vacuum. This procedure can be summarized by stating that the excitations over the initial state can only have positive~energy. 

\begin{figure}
\includegraphics[scale=.5]{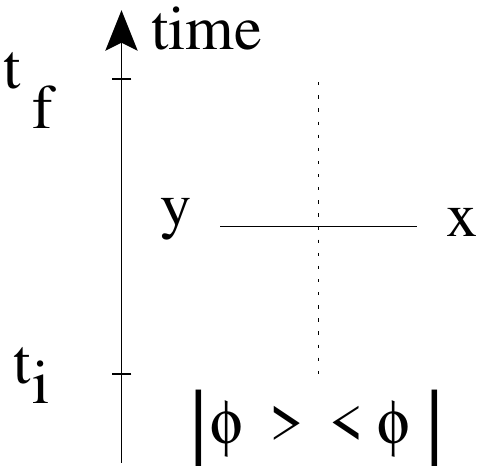}\hskip2.6cm\includegraphics[scale=.5]{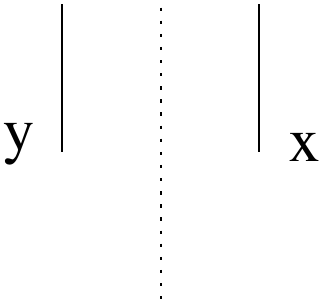}

\hskip1.4cm(a)\hskip4cm(b)
\caption{A solid line stands for the propagator in a CTP Feynman graph. The~vertical dotted line separates the ket  and the bra sectors where the field variable $\phi_+$ and $\phi_-$ are used, respectively. The~time runs upward. (\textbf{a}): The Wightman function $D_{-+}(x-y)$ connects the bra and the ket components. (\textbf{b}): An alternative representation of the Wightman function where the lines follow the world lines of the excitations until the final time when the trace operation carried out in \eq{effgenfunct} connect them. These excited states are~on-shell.}\label{wightmannf}
\end{figure}

To assess the importance of the positivity of the excitation energies, let us consider the off-diagonal block of the propagator, 
\be
iD_{-+}(x-y)=\sum_n\la n|\phi(y)\rho(t_i)\phi(x)|n\ra,
\ee
where the trace is obtained by summing over the stationary states of the full closed dynamics. The~contributions come from excited bra and ke states, $|\phi(x)\ra$ and $\le\phi(y)|$, respectively. Hence this block of the propagator is non-vanishing only for positive energies states $|n\ra$, c.f. Figure \ref{wightmannf} and Equation \eq{invfrpr}. A~similar argument applies to $D_{+-}(x-y)$ which is non-vanishing for negative energies. It is easy to see that this structure is inherited by the higher order Green function; in addition, $\Tr[T[\prod_j\phi(y_j)]\rho T[\prod_k\phi(x_k)]^*\rho]$ is vanishing when the total energy flowing from the legs $y_j$ to $x_k$ is~negative.

\begin{figure}
\includegraphics[scale=.5]{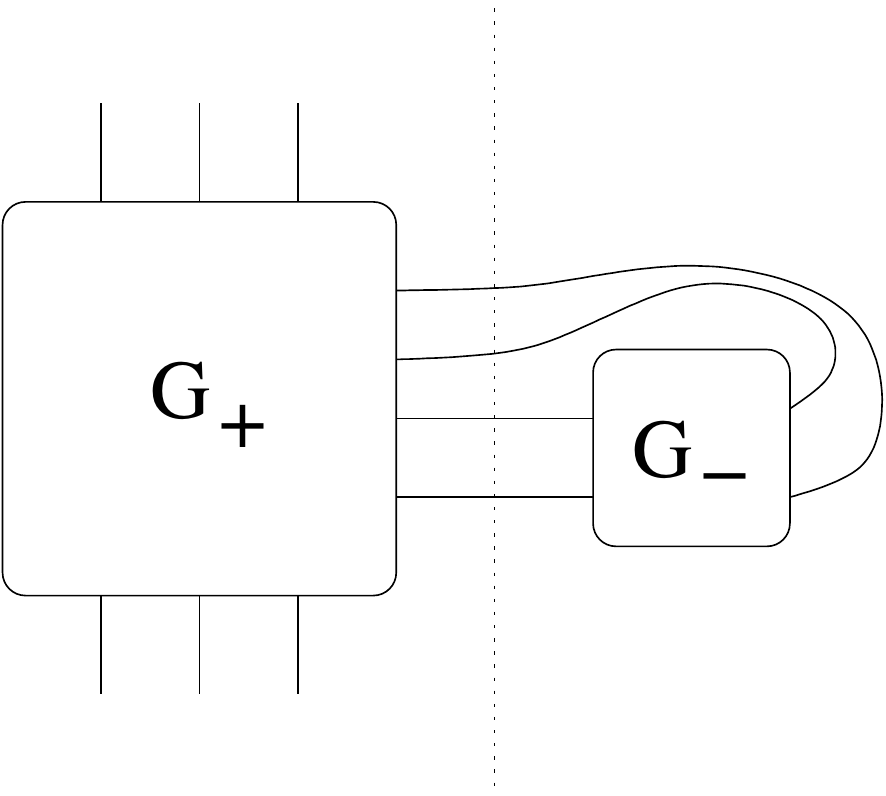}\hskip1.6cm\includegraphics[scale=.5]{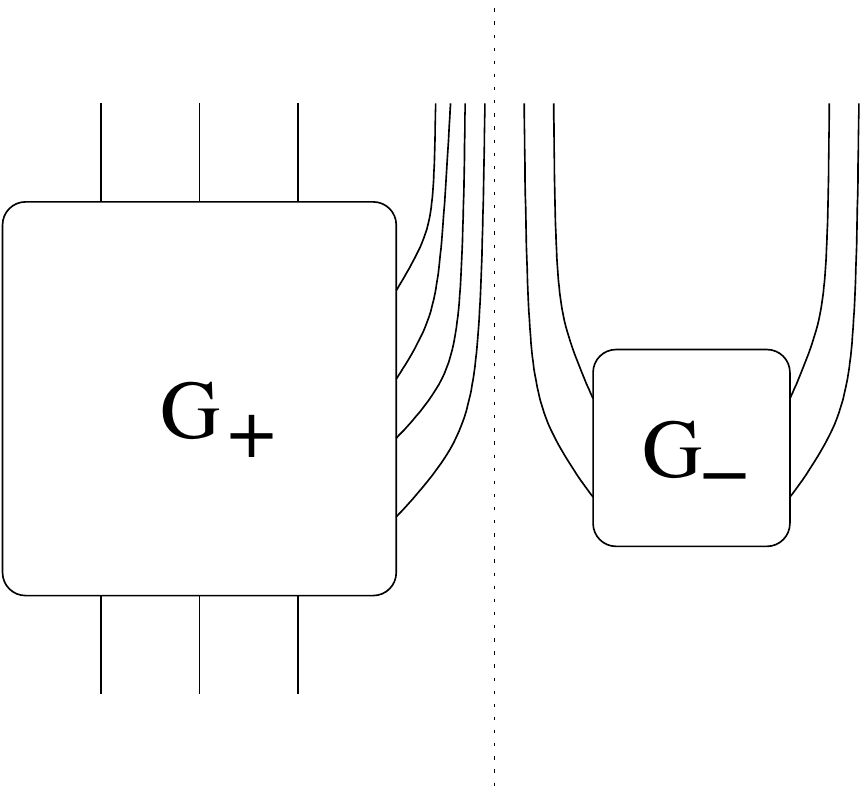}

\hskip2.5cm(a)\hskip5.5cm(b)
\caption{An inhomogeneous Feynman graph contributing to a sixth order Green function of the field $\phi_+$. The~subgraphs $G_+$ and $G_-$ belong to ket and bra excitations, respectively. (\textbf{a}): The sum of the energies flowing into $G_-$ is vanishing hence at least on of the line carries negative energy and the graph is vanishing. (\textbf{b}): The alternative representation with the lines indicating the excitations contributing to the trace. This process is suppressed if the energy is vanishing at the initial~time.}\label{subgrf}
\end{figure}

Such a restriction leads to a remarkable simplification of the interactive Green functions~\cite{effthe}. The~CTP Feynman graphs can be grouped into three classes: the homogeneous graphs are where all external and internal lines belong to the same CTP copy. The~external legs correspond to the same CTP copy but there are end points of the internal lines land at both copies in inhomogeneous graphs. Finally, the genuine CTP graphs have external legs attached to both copies. The~homogeneous graphs are identical to the their STP counterparts and the genuine CTP graphs represent processes which generate excitations at the final time. The~interesting question is whether the inhomogeneous graphs differ in STP and CTP. The~answer to this question is negative if the excitations have positive energy, cf. Figure~\ref{subgrf}.

\unskip

\subsection{Open Interaction~Vertices}
The action $S[\phi_+,\phi_-]$ of an open dynamics couples the field variables corresponding to the bra and the ket. We have seen that such a coupling must be non-local in time and the simplest classification of non-local action is the cluster expansion. The~lowest order which incorporates the condition of positive energy excitations is the bilocal level,
\bea\label{bact}
S[\hphi]&=&\hf\int dx\hphi(x)\hD^{-1}\hphi(x)-\int dx[U(\phi_+(x))-U^*(\phi_-(x))]\nn
&&-\int dxdyV(x-y,\phi_-(x),\phi_+(y)).
\eea

The kernel of the first term is the inverse of the massless free propagator \eq{invfrpr} and the conservative interaction is described by a local potential,
\be
U(\phi)=\sum_{0\le n\le N}\frac{g_n}{n!}\phi^n.
\ee

The open interactions are represented by a bilocal potential,
\be\label{bicolpot}
V(x-y,\phi_-,\phi_+)=\tilde\Theta(x-y)\sum_{0\le n_++n_-\le N}\frac{h_{n_-,n_+}}{n_-!n_+!}\phi^{n_-}_-\phi^{n_+}_+,
\ee
where $\tilde\Theta(x-y)=\tilde\Theta(x^0-y^0)\delta(\v{x}-\v{y})$. Though~one could in principle extend the ansatz by allowing a non-trivial dependence in $x^\mu-y^\mu$ such a restriction is needed for the invariance with respect to the orthochronuos Lorentz group, the~subgroup of the full Lorentz group which preserves the direction of the time. The~symmetry with respect to time reversal of the open system dynamics is broken by the initial condition of the environment. The~potentials are complex, $g_n=g_{nr}+ig_{ni}$, $h_{mn}=h_{mnr}+ih_{mni}$, the~imaginary part is generated by the intermediate states of the Feynman diagrams  according to the optical theorem and the time reversal invariance of the full closed dynamics requires $h_{mmr}=0$. A~closed theory is obtained by the choice $g_{2i}=-\epsilon$, $h_{11i}=2\epsilon$ and vanishing higher orders, $h_{n_-,n_+}=0$ for $n_\pm>1$.

We use below the truncation with $N=4$,
\bea
U(\phi)&=&\frac{g_2}2\phi^2_++\frac{g_4}{4!}\phi_+^4,\nn
V(x-x',\phi_-,\phi_+)&=&\tilde\Theta(x-x')\left(ih_{11i}\phi_-\phi_++i\frac{h_{22i}}4\phi^2_-\phi^2_+\right).
\eea

One has in principle $\ord{\phi_-\phi_+^3}$ and $\ord{\phi^3_-\phi_+}$ vertices but their evolution is of fourth order in the cluster expansion hence they are ignored in our second order truncation scheme. In~the absence of energy conservation, the stability follows from the finiteness of the path integral $\mr{Im}~S[\hphi]>0$ which amounts to $g_{4i}<12|h_{22i}|$. Since we assume the triviality of the saddle point for the blocking the condition $g_{2i}<0$ has to be imposed, too.

\subsection{UV-IR Entanglement and~Decoherence}
A distinguished feature of an open quantum system is that the system-environment entanglement renders the system state mixed. A~mixed state consists of several pure states, the~eigenstates of the density matrix, corresponding to a probability which is given by the corresponding eigenvalue. It is crucial to note that different pure states do not enter into interference with each other in the expectation value of observables~\cite{equilibrium}. Such a restriction of the coherence is usually called decoherence~\cite{zehd,zurekd}, defined roughly as the suppression of the off-diagonal elements of the density matrix. Naturally, such a definition depends on the basis where the off-diagonal elements are taken. The~decoherence is displayed below in the field diagonal basis by the help of the influence functional~\cite{feynman}.

It is enlightening to employ the parameterization $\phi_\pm=\phi\pm\phi_d/2$, by~interpreting $\phi$ and $\phi_d$ as the classical field and the quantum fluctuations, respectively. This comes from the observation that the decoherence in the field's diagonal basis arising in the classical limit suppresses $\phi_d$ and the expectation value of any functional of $\phi_\pm$ agrees with that of $\phi$. 

The quantum fluctuation $\phi_d$ is suppressed by the real part of the exponent in the path integral \eq{genfpi}. Hence the generic $\phi_d$ suppression, present for $\phi=0$, is driven by its $\phi$-independent part,
\be
-\mr{Im}S_{|\phi=0}=\int\frac{dp}{(2\pi)^4}\left[\phi_{d,-p}\frac{g_{2i}-h_{11i}\Theta_{p^0}}4\phi_{d,p}+(\phi_d^2)_{-p}\frac{g_{4i}+\frac{h_{22i}}4\Theta_{p^0}}{12}(\phi^2_d)_p\right],
\ee
where $(\phi_d^2)_p$ denotes the Fourier transform of $\phi^2_d(x)$. The~finiteness of the life-time of the quasi-particles created by $\phi(x)$ and $\phi^2(x)$ in the closed dynamics is encoded by $g_{2i}$ and $g_{4i}$ in a frequency independent manner. Though~these parameters appear in the closed part of the action they represent both closed and open interactions, the~finite life-time formed in a closed system and the leaking of the quasi-particle state into the environment via the open interaction channels. The~environment-induced decoherence appears exclusively as the modification of the quantum fluctuations owing to the interaction with the environment at positive energies, described by the parameters $h_{11i}$ and $h_{22i}$. It is remarkable that decoherence may turn into recoherence depending on the sign of the parameters, the~latter indicating the emergence of coherent structures due to the~environment. 

The $\phi_d$-dependent part of $\mr{Im}S[\phi,\phi_d]$ for a given $\phi(x)$ describes the decoherence and recoherence of the classical field configuration $\phi(x)$. Though~the joint dynamics of both $\phi$ and $\phi_d$ is stable as long as $g_{4i}<12|h_{22i}|$ the recoherence of a particular classical field becomes strong for 
\be\label{intlphs}
-\frac{h_{22i}}4<g_{4i}<12|h_{22i}|.
\ee

\section{Gliding~Cutoff}\label{btrs}
The central point of this work, the~need for treating the mixed components of the vacuum state of a cutoff theory is discussed with the help of the functional renormalization group method within the CTP formalism, introduced in this~section.

\subsection{Euclidean Field Theory at Thermal~Equilibrium}
Lowering the cutoff is the simplest to cast in terms of the partition function of an Euclidean quantum field theory at finite temperature, written in a path integral form,
\be\label{epartf}
\Tr[e^{-\beta H_k}]=\int D[\phi]e^{-S_k[\phi]},
\ee
where the field is periodic in time with period length $\beta$ and the regularization procedure is considered a part of the action, $S_k[\phi]$, $k$ denoting the gliding cutoff. The~right hand side is considered to be a partition function of a $d$-dimensional classical statistical physical system with a Hamiltonian $S_k[\phi]$ at unit temperature. The~cutoff should be introduced only for the spatial components of the momentum to preserve the~temperature.

The blocking of the bare dynamics consists of the decrease of the UV cutoff, $k\to k-\Delta k$ with $\Delta k=(\delta_-+\delta_+)k$ and the splitting the field variable into the sum $\phi\to\phi+\varphi$, where $\phi$ and $\varphi$ contains the IR ($|\v{p}|<k(1-\delta_-)$) and the UV ($k(1-\delta_-)<|\v{p}|<k(1+\delta_+)$) modes, respectively. The~blocked action of the thinned theory is found by integrating over the UV field~\cite{wh},
\be\label{eblor}
e^{-S_{k-\Delta k}[\phi]}=\int D[\varphi]e^{-S_k[\phi+\varphi]}.
\ee

One should in principle follow the cutoff-dependence of the generator functional for the connected Green functions,
\be
e^{W_k[j]}=\int D[\phi]e^{-S_k[\phi]+\int dxj(x)\phi(x)},
\ee
to keep track of the cutoff-dependence of the dynamics. However, the presence of the IR field on both side of the blocking relation \eq{eblor} allows us to follow the evolution of the dynamics in the blocked action directly. The~initial condition for the renormalized trajectory is the bare action at the initial  UV cutoff, $\Lambda$.

\subsection{Real Time Dynamics of Quantum Field~Theories}
The realization that the change of the cutoff can be treated in a similar manner in classical and quantum statistical physics had a strong impact on our way to handle many body systems. It arose from interpreting the $S_k[\phi]$ in \eq{eblor} either as the potential energy of a classical field theory in $d+1$ dimensions or as the action of an Euclidean $d$-dimensional quantum field theory. However there is a fundamental difference between the classical and the quantum dynamics, namely the entanglement, which forces us to follow a different route in the case of quantum~systems.

{We continue with an isolated, closed dynamics with the initial value $\Lambda$ of the cutoff.} The blocked action \eq{epartf} can be used in thermal equilibrium to obtain the reduced density matrix and the canonical partition function of the IR modes. However the IR-UV entanglement creates a problem when the real time effective dynamics is sought. The~traditional use of \eq{eblor} is to find the usual Green functions for the IR field, generated by 
\be\label{rtgenf}
\la0_{\Lambda}|U(t_f,t_i;j)|0_{\Lambda}\ra=\int D[\phi]e^{iS_{\Lambda}[\phi]+i\int dxj(x)\phi(x)}
\ee
where $|0_{\Lambda}\ra$ denotes the vacuum of the closed cutoff theory with the initial cutoff $\Lambda$. The~problem with this expression is that it corresponds to a transition amplitude between pure states while the elimination of a dynamical degree of freedom generates a mixed state. In~other words, the~blocking takes us beyond the traditional STP formalism of quantum field theory and forces us to use the reduced density matrix to represent the state of the retained degrees of freedom. One can naturally construct the reduced density matrixes by convoluting Green functions with different final states with the density matrix of the full system. Rather than following such an involved scheme we turn to the CTP formalism where these final state sums are already build in to streamline the calculation and to have more transparent~equations.

\subsection{Changing the Cutoff in Open Quantum~Systems}
The generalization of the blocking \eq{eblor} for CTP follows the steps of ref.~\cite{wh}, by~starting with
\be\label{blrel}
e^{iS_{k-\dk}(\hphi)}=\int D[\hat\varphi]e^{iS_k[\hphi+\hat\varphi]}
\ee
and the continuing with the one-loop approximation,
\be\label{fdeveq}
e^{iS_{k-\dk}(\hphi)}=\int D[\hat\varphi]e^{iS_k[\hphi+\hat\varphi_0]+\frac{i}2\hat\varphi\fdd{S_k[\hphi+\hat\varphi_0]}{\hat\varphi}{\hat\varphi}\hat\varphi+\ord{\dk^2}}.
\ee

The unexpected strength of this procedure is the emergence of a one-loop equation which is exact. In~fact, the~limit of infinitesimal blocking $\delta_-+\delta_+=\dk\to0$ suppresses the higher loops contributions. To~simplify the resulting evolution equation one assumes that the saddle point is trivial, $\hat\varphi_0=0$. One arrives in this manner at the Fresnel integral
\be
e^{i[S_{k-\dk}(\hphi)-S_k(\hphi)]}=\int D[\hat\varphi]e^{\frac{i}2\hat\varphi\fdd{S_k[\hphi]}{\hat\varphi}{\hat\varphi}\hat\varphi}
\ee
which yields the { CTP form of the Wegner-Houghton} equation
\be\label{eveq}
\dot S[\hphi]=-i\frac{k}2\Tr\ln\left[\fdd{S}{\hphi}{\hphi}\right],
\ee
where the trace is over the UV field space and the dot stands for the derivative with respect to $\tau=\ln k/\Lambda$. {The solution of the evolution equation is rendered unique by specifying the initial condition, the~bare action at $k=\Lambda$.} The compactness Equation \eq{blrel} hides the an essential element of the blocking in quantum systems: {We seek the reduced density matrix \eq{densm} for the IR modes hence these are handled in the OTP formalism by the help of the blocked action. The~elimination of the UV modes and the~execution of the partial trace in \eq{densm}, is carried out in the CTP formalism.} {\em The blocking is the placing of the modes to be eliminated from the OTP to the CTP scheme.} Without the infinitesimal off-diagonal term of \eq{invfrpr} in the free UV action the IR action remains additive and \eq{blrel} represents the product of two independent STP~amplitudes.

To make the solution of this evolution equation feasible we project it onto the functional space of the bilocal action \eq{bact}. This step transforms the evolution Equation \eq{eveq} into a set of coupled differential equations for the running parameters of the blocked action. These parameters are defined by evaluating the blocked action on a family of IR field configurations, $\phi_s(x)$, called subtraction point. The~parameters of a cutoff theory characterize the physics at the cutoff scale, hence the subtraction point should be placed close to the gliding cutoff. The~imaginary time theories are free of mass-shell singularities and one customarily places their subtraction point at the IR end, at~a homogeneous field configuration $\phi_x(x)=\Phi$, by~hoping that the truncated gradient expansion can still reproduce the desired dynamics around the cutoff scale. The~real time dynamics is dominated by the propagating quasi-particle modes hence the subtraction point should be placed into their kinematical region~\cite{rgmink}. Thus, the evolution equation is evaluated at the subtraction point, defined by the IR field configuration $\phi^{(s)}_{\omega,\v{p}}=\Phi_\pm(2\pi)^4\delta(\v{p})\rho_\omega$ where $\phi_q=\int dxe^{iqx}\phi(x)$ and
\be\label{subtrp}
\rho_\omega=\frac\eta{2\pi}\left[\frac1{(\omega-\omega_s)^2+\eta^2}+\frac1{(\omega+\omega_s)^2+\eta^2}\right],
\ee
$\omega_s=c_s\sqrt{k^2+g_{2r}}$, $c_s\ge1$ being a cutoff-independent dimensionless parameter of the subtraction scheme. The~$\eta$ parameter introduces a regular wave packet in time and a monochromatic subtraction point in the limit $\eta\to0$.

\subsection{Evolution~Equation}\label{eveqs}
The contribution of the closed local part to the left hand side of the evolution equation at the subtraction points, 
\be
\int dtU(\Phi\rho(t))=\sum_n\frac{g_nu_n}{n!}\Phi^n,
\ee
yields
\be
u_n=\left(\frac{i\eta}{2\pi}\right)^n\int dte^{-n\eta|t|}\left(\frac{e^{i\omega _rt}}{\omega_s+i\eta}-\frac{e^{-i\omega _rt}}{\omega_s-i\eta}\right)^2.
\ee

The open part contributes by
\be
\int dtdt'V(t-t',\phi_-(t),\phi_+(t'))=\sum_{n_\pm=0}^\infty\frac{h_{n_-,n_+}u_{n_-,n_+}}{n_-!n_+!}\Phi^{n_-}_-\Phi^{n_+}_+,
\ee
with
\bea
u_{n_-,n_+}&=&\left(\frac{i\eta}{2\pi}\right)^{n_-+n_+}\int dtdt'\int_0^\infty\frac{d\omega}{2\pi}e^{-i(t-t')\omega-(n_-+n_+)\eta(|t|+|t'|)}\nn
&&\times\left(\frac{e^{i\omega _rt}}{\omega_s+i\eta}-\frac{e^{-i\omega _rt}}{\omega_s-i\eta}\right)^{n_-}\left(\frac{e^{i\omega _rt'}}{\omega_s+i\eta}-\frac{e^{-i\omega _rt'}}{\omega_s-i\eta}\right)^{n_+}.
\eea

The right hand side of the evolution equation contains the second functional derivative,
\be
\fdd{S}{\hphi_p}{\hphi_q}=\hD^{-1}_p\delta_{p,q}-\hat\Sigma_{p,q}
\ee
with $\delta_{p,q}=(2\pi)^4\delta(p-q)$. The~inverse propagator,
\be
\hD_p^{-1}=\begin{pmatrix}p^2-g_2+i\epsilon&-\frac{i}2h_{11i}[1-\sign(p^0)]\cr-\frac{i}2h_{11i}[1+\sign(p^0)]&-p^2+g_2^*+i\epsilon
\end{pmatrix},
\ee
yields the propagator
\be\label{drprop}
\hD_p=\begin{pmatrix}\frac1{p^2-g_2}&-\frac{ih_{11i}}{(p^2-g_{2r})^2+g_{2i}^2}\Theta_{-p^0}\cr-\frac{ih_{11i}}{(p^2-g_{2r})^2+g_{2i}^2}\Theta_{p^0})&-\frac1{p^2-g^*_2}\end{pmatrix},
\ee
in the absence of the IR subtraction field. It contains the Feynman propagator with complex mass in the diagonal and the corresponding Lorentz-spread mass-shell condition in the off-diagonal matrix elements. The~self energy is
\bea
\Sigma_{(+,p)(+,p')}&=&\delta_{\v{p},\v{p}'}\hf I_{p'^0-p^0}(g_4\Phi_+^2+\Theta_{p'^0-p^0}h_{22}\Phi_-^2),\nn
\Sigma_{(-,p)(-,p')}&=&\delta_{\v{p},\v{p}'}\hf I_{p^0-p'^0}(\Theta_{p^0-p'^0}h_{22}\Phi_+^2-g^*_4\Phi_-^2),\nn
\Sigma_{(-,p)(+,p')}&=&\delta_{\v{p},\v{p}'}J_{-p^0,-p'^0}h_{22}\Phi_+\Phi_-,\nn
\Sigma_{(+,p)(-,p')}&=&\delta_{\v{p},\v{p}'}J_{p'^0,p^0}h_{22}\Phi_+\Phi_-,
\eea
with
\bea
I_\omega&=&\int\frac{d\omega'}{2\pi}\rho_{\omega-\omega'}\rho_{\omega'},\nn
J_{\omega,\omega'}&=&\int\frac{d\omega''}{2\pi}\Theta_{\omega''}\rho_{\omega''+\omega}\rho_{\omega''+\omega'}.
\eea

\begin{figure}
\includegraphics[scale=.5]{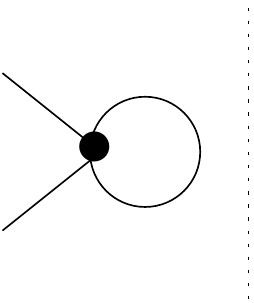}\hskip1.6cm
\includegraphics[scale=.5]{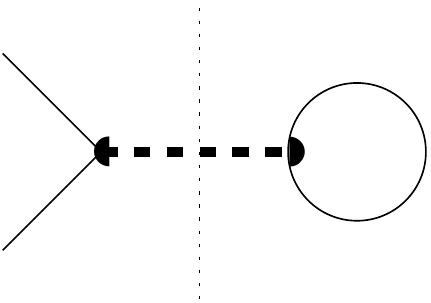}\hskip1.6cm
\includegraphics[scale=.5]{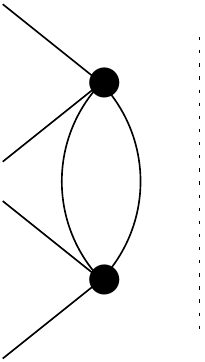}
\caption{The Feynman graphs contributing to the STP parameters. The~horizontal dashed line connects the two clusters of the bi-local vertices which couple the bra and the~ket.}\label{stpbff}
\end{figure}

\begin{figure}
\includegraphics[scale=.5]{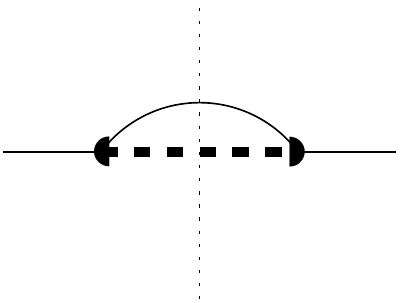}\hskip.6cm
\includegraphics[scale=.5]{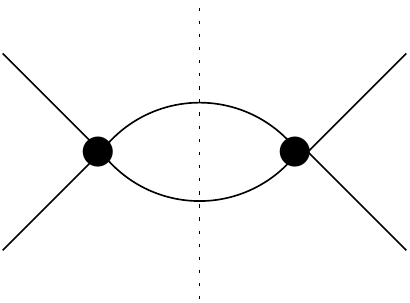}\hskip.6cm
\includegraphics[scale=.5]{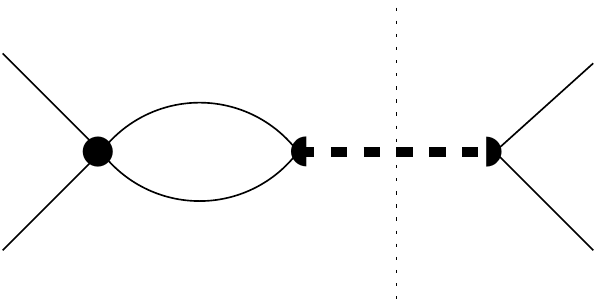}
\caption{The Feynman graphs contributing to the CTP parameters. The~time inverted version of the last graph enters, as~well.}\label{ctpbff}
\end{figure}

The Neumann expansion of the right hand side of the evolution equation in the self energy,
\be\label{expevqe}
\dot S=-i\frac{k}2\left(\Tr[\ln\hD^{-1}]-\sum_{n=1}^\infty\frac1n\Tr[(\hD\hat\Sigma)^n]\right),
\ee
is sufficient up to the quadratic order and the identification of the coefficients of the terms $\ord{\Phi_\pm^n}$, $n\le4$ produces the beta functions, the~derivatives of the parameters with respect to $t$,
\bea\label{evoleqode}
\dot g_2&=&Ag_4,\nn
\dot g_4&=&Bg^2_4,\nn
\dot h_{11i}&=&Ch_{11i}h_{22i},\nn
\dot h_{22i}&=&Dg^*_4g_4h^2_{11i}+Eg_4h_{22i}+E^*g^*_4h_{22i},
\eea
cf. Figures~\ref{stpbff} and \ref{ctpbff} where the $g_2$-dependent coefficients are
\bea\label{loopint}
A&=&-i\frac{k^3I_0}{4\pi^2u_2}\int\frac{d\omega}{2\pi}D^{(k)}_{++\omega},\nn
B&=&-i\frac{3k^3}{4\pi^2u_4}\int\frac{d\omega}{2\pi}\frac{d\omega'}{2\pi}I_{\omega-\omega'}^2D^{(k)}_{++\omega}D^{(k)}_{++\omega'},\nn
C&=&-i\frac{k^3}{4\pi^2u_{1,1}}\int\frac{d\omega}{2\pi}d^{(k)}_\omega(\Theta_{-\omega}J_{-\omega,-\omega}+\Theta_{\omega}J_{\omega,\omega}),\nn
D&=&\frac{k^3}{4\pi^2u_{2,2}}\int\frac{d\omega}{2\pi}\frac{d\omega'}{2\pi}d^{(k)2}_\omega \Theta_{-\omega}\Theta_{\omega'}I_{\omega-\omega'}^2,\nn
E&=&-i\frac{k^3}{8\pi^2u_{2,2}}\int\frac{d\omega}{2\pi}\frac{d\omega'}{2\pi}D^{(k)}_{++\omega}D^{(k)}_{++\omega'}I^2_{\omega-\omega'}\Theta_{\omega-\omega'},
\eea
with $\hD^{(k)}_\omega=\hD^{(k)}_{\omega,k\v{n}}$ and $\v{n}^2=1$, $d^{(k)}_\omega=-i/[(\omega^2-k^2-g_{2r})^2+g_{2i}^2]$. The~higher than two-cluster contributions arising from the Neumann expansion have been ignored in deriving the beta functions. The~integrations can been carried out~analytically.

\unskip

It is remarkable that the STP beta functions contain no open parameters; in~other words, the STP parameters evolve as in the traditional STP formalism of Quantum Field Theory. This follows immediately from the equivalence of the inhomogeneous CTP and the STP graphs, mentioned in Section~\ref{insts}.

\subsection{Separatrices and Phase~Transition}\label{uvirscs}
A quantum phase transition corresponds to a separatrix of the renormalization group flow, indicating that small modifications of the theory in the UV lead to large changes in the IR. One cannot strictly establish a phase transition with the help of a truncated renormalization group flow but it is reasonable to assume that while stable flows of an appropriately truncated flow represent a good approximation of the exact case this does not hold for trajectories with IR singularity. In~fact, the latter never occurs for the exact flow and suggests that important parameters are missed due to the limited ansatz space for the blocked action. Hence the trajectory on the border of the stable IR flow indicates a separatrix of the exact~solution.

One can gain some qualitative insight into the scaling laws of the closed parameters by writing the first two equation of \eq{evoleqode} as a single equation for $x=g_2$,
\be\label{analcp}
\ddot x=-\nu_x\dot x+\xi\dot x^2,
\ee
describing the complex trajectory of a one-dimensional damped motion with
\bea
\nu_x&=&-\frac{\partial_\tau A}A,\nn
\xi&=&\frac{\partial_xA+B}A
\eea
in terms of the complex beta function parameters $A(\tau,x)$ and $B(\tau,x)$. The~quartic coupling is found by $g_4=\dot x/A$. It is instructive first to inspect this equation in simpler~cases. 

In an $O(4)$ invariant Euclidean field theory the parameters are real and one finds at the subtraction point $\phi^{(s)}(x)=0$
\bea
A_E&=&-\frac{k^4}{16\pi^2\omega_k^2},\nn
B_E&=&\frac{3k^4}{16\pi^2\omega_k^4},
\eea
and
\bea
\nu_{xE}&=&-2\frac{k^2+2g_2}{k^2+g_2},\nn
\xi_E&=&-\frac4{k^2+g_2}.
\eea

The renormalized trajectory starting with the initial conditions $g_2(\Lambda)\ll k^2_{in}$, $\dot g_2(\Lambda)=Ag_4(\Lambda)$ stretches toward negative $\tau$. In~the UV scaling regime, $k^2>|g_2(k)|$, $\nu_x\approx-2$, $\xi\approx0$, and~the evolution starting with positive velocity (towards decreasing $t$!) is slowed down by the friction. In~the IR regime, $k^2<|g_2(k)|$,  $\nu_x\approx-4$, $\xi\approx-4/g_2$, both the friction and the $\ord{\dot x^2}$ term continue to damp the evolution. However, for sufficiently negative $g_2(\Lambda)$ the damped increase of $g_2$ may be slow enough to reach $g_2(k)=-k^2$ at $k=k_{sp}>0$. As~this crossover is approached $B$ diverges sending $g_4$ to zero and an IR singularity is~generated.

In the real time theory with sharp momentum cutoff at the subtraction point $\phi^{(s)}(x)=0$ one finds
\bea
A_{M,0}&=&-\frac{k^3}{8\pi^2\omega_k},\nn
B_{M,0}&=&\frac{3k^3}{16\pi^2\omega_k^3},
\eea
and
\bea
\nu_{x,M,0}&=&-\frac{2k^2+3g_2}{k^2+g_2},\nn
\xi_{M,0}&=&-\frac2{k^2+g_2}.
\eea

The parameters remain real and the symmetry broken phase is recovered in a qualitatively similar manner as in the imaginary time~case.

The evolution of the theory in Minkowski space-time defined by a plane wave subtraction point \eq{subtrp} with $\eta\to0$ at the threshold, $c_s=1$, is driven by
\bea
A_{M,1}&=&-\frac{k^3}{4\pi^2\omega_k},\nn
B_{M,1}&=&\frac{6k^3}{\pi^2\omega^2_k}\left(1-i\frac{\omega_k}{2g_{2i}}\right),
\eea
where
\bea
\nu_{x,M,1}&=&-\frac{2k^2+3g_2}{k^2+g_2},\nn
\xi_{M,1}&=&-\frac{49}{2(k^2+g_2)}+\frac{12i}{g_{21}},
\eea
and the symmetry broken phase is recovered as~well. The~complex trajectory may avoid the singularity by having $g_{2i}(k_{sp})\ne0$. A~short enough finite life-time of the quasi particles may weaken the crossover singularity and render the simple ansatz for the action applicable in the IR region within the symmetry broken~phase. 

The evolution of the open parameters can be read off from the scale-dependence of $y=\ln h_{11j}$ satisfying the equation
\be\label{analp}
\ddot y=-\nu_y\dot y-U'(y)
\ee
corresponding to a one-dimensional particle of unit mass, moving under the influence of a friction force with Newton constant
\be\label{fricc}
\nu_y=-\frac{\dot C}C-(Eg_4+E^*g_4^*),
\ee
and a potential
\be\label{pote}
U(y)=-\frac{CD}2g_4^*g_4e^{2y}.
\ee 

The real beta function parameters are $\tau$-dependent, $C(\tau,x(\tau)),D(\tau,x(\tau))$ and $E(\tau,x(\tau))$. The~other open parameter is given by $h_{22i}=\dot y/C$. The~trajectory starting in the vicinity of the Gaussian fixed point where  $C,D,E<0$ with $y(0)\approx-\infty$ and $\dot y(0)\approx0$. The~dominant scale-dependence of $C$ and $D$ with finite $\eta$ is $e^{3\tau}$, which weakens the potential but keeps the friction stable and approximately scale invariant. The~coordinate $y$ rolls down on the potential in the positive direction reflecting the irreversible accumulation of the system-environment entanglement during the change of the cutoff. The~entanglement is weak for short lived quasi particle excitations hence $C$ and $D$ decreases with increasing $-g_{2i}(\Lambda)$. Thus, the exponentially fast decreasing potential cannot destabilize the evolution for large enough $-g_{2i}(\Lambda)$. However, the lowering of $-g_{2i}(\Lambda)$ strengthens the potential and the exponentially steep potential may make the trajectory divergent at finite scale, generating a separatrix for the flow of the open parameters. The~evolution in the UV direction makes the instability stronger since $\nu<0$.

\section{Renormalization Group~Flow}\label{rgflows}
The issues we intend to comment on or clarify by the numerical integration of the evolution equations are (1) the phase structure of the theory, (2) the closed bare theory limit, (3) the relevance of the open parameters of the action and (4) the renormalizability. We do this by exploring the renormalization group flow restricted to initial conditions in the vicinity of the Gaussian fixed~point.

\subsection{Phase~Structure}
The closed parameters evolve independently from the open channels hence the usual phase transition between the $Z_2$ symmetrical and the spontaneous broken phases takes place at the same place as in the closed theory. The~singularity at $\tau=\tau_{sp}$ is a spinodal instability indicating that the vacuum is in the symmetry broken phase~\cite{more,caillol,palaez,tree,vincent}. The~IR singularity of the open channels indicates that the theory may undergo a phase transition where the system-environment interactions increases abruptly for the IR~modes.

\begin{figure}
\includegraphics[scale=1]{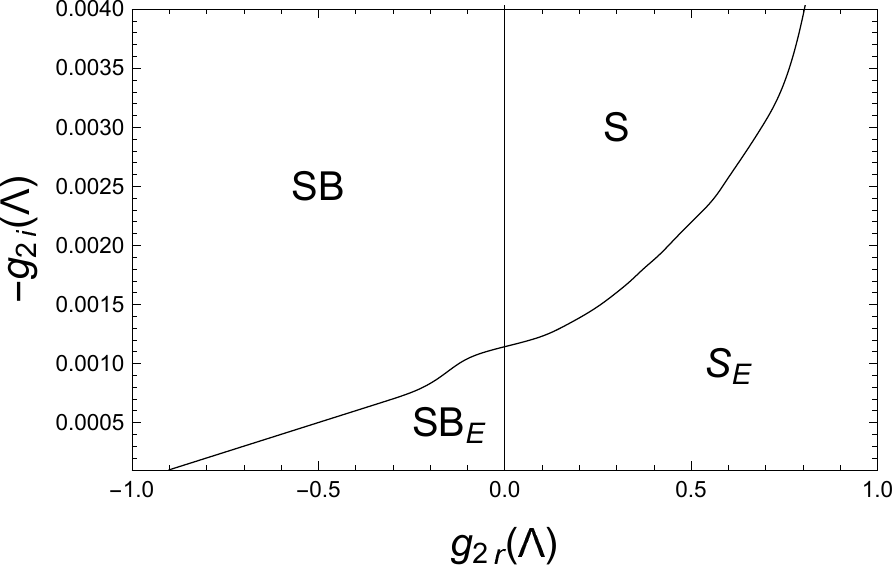}
\caption{The phase structrure on the complex $g_2(\Lambda)$ plane at $g_4(\Lambda)=0.1$, $h_{11}(\Lambda)=-i 2 g_{11i}(\Lambda)$, $h_{22}(\Lambda)=0$ and $\eta=1$, S: symmetrical phase, SB: symmetry broken phase and the entengled phases with the subscript $E$.}\label{phstrf}
\end{figure}

Such a phase structure is borne out by the integration of the evolution equations \eq{evoleqode}. The~four phases are shown on the complex $g_2(\Lambda)$ plane of Figure~\ref{phstrf}. The~spontaneous breakdown of the $Z_2$ symmetry is indicated within the framework of the local potential approximation by the divergence of the propagator, $k^2+g_2(k)\to0$, followed by a spinodal instability as the cutoff is lowered. The~transition between the symmetric and the symmetry broken phase is a slightly right bended vertical line. The~phases with regular or divergent $h_{22i}$ are separated by the curve which increases with $g_{2r}$. 

When the quasi-particles are stable enough to interact with the environment, the~theories below the curve, the~evolution drives $h_{22i}\to\infty$ at finite cutoff. The~large positive $h_{22i}$ turns on the interactions with the environment indicating a large amount of system-environment entanglement generated by the lowering of the cutoff. Hence the phases under the curve are called symmetric or symmetry broken entangled phases. The~exact renormalized trajectories are regular in the IR direction and the singularity in the entangled phase indicates the insufficiency of our ansatz to give account of the increased amount of entanglement. A~safe conclusion, point (iii) of the introduction, is that the long range macroscopic dynamics turns suddenly non-perturbative in terms of the microscopical quasi-particles at the curve separating the upper and the lower phases. With~an improved ansatz for the action which allows us to penetrate the entangled phase the lower end of the separation of the phases $S_E$ and $SB_E$ should become visible since the STP parameters evolve independently from the open~interactions.
The Lorentzian width $\eta$ of the subtraction point controls the energy interval around the $\omega_s$ where the contributions to the evolution of the running coupling constants are read off from the right hand side of Equation \eq{eveq} and smaller width makes the propagating quasi-particles dominate. The~general trend is that the entangled phase shrinks with the increase of $\eta$, we need propagating quasi-particles to pick up the system-environment entanglement. The~decrease of $\eta$ leads ultimately to numerical instabilities, c.f. Section~\ref{closedts}. Among~the initial conditions with reliable solution we found no example that the closed limit $-g_{2,i}(\Lambda)=h_{11i}(\Lambda)/2\to0$, cf. the inverse propagator \eq{invfrpr}, with~$g_{4i}=h_{22i}=0$ would avoid the entangled~phase.

\begin{figure}
\includegraphics[scale=.7]{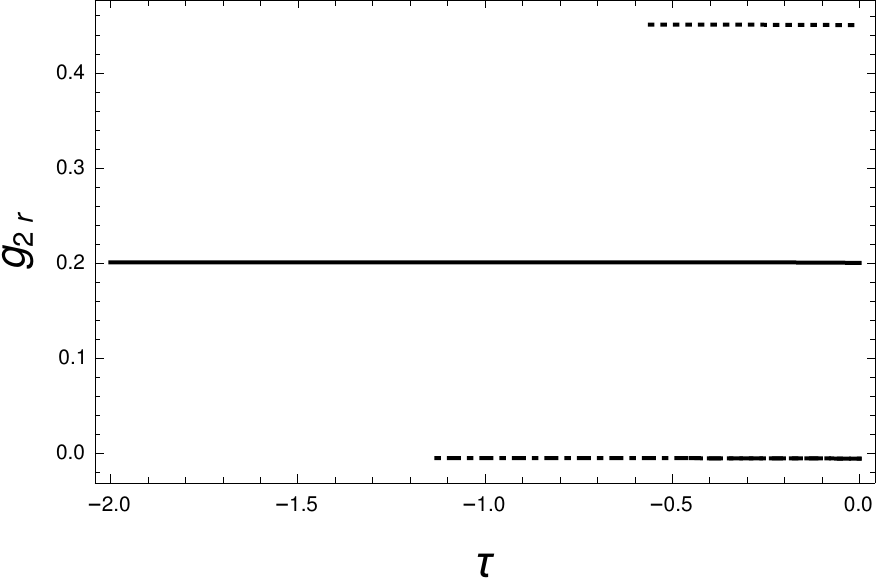}\hskip.5cm\includegraphics[scale=.7]{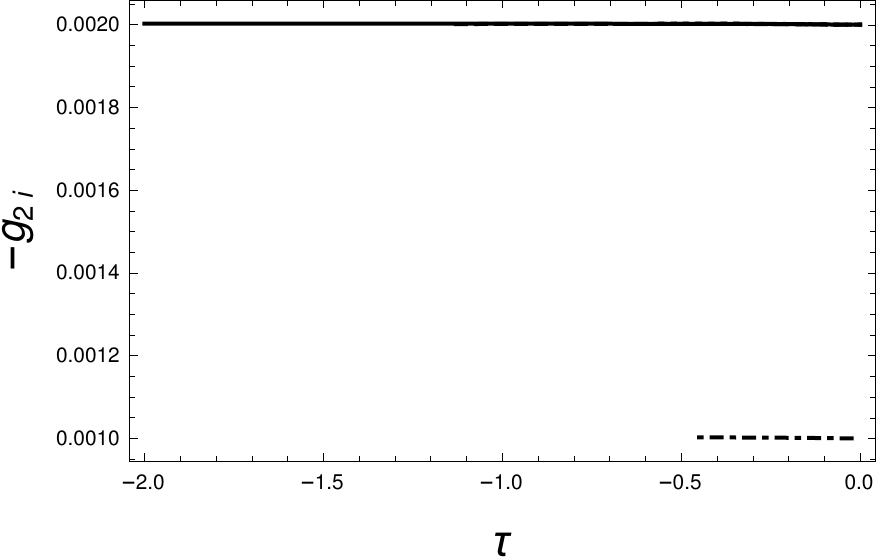}

\includegraphics[scale=.7]{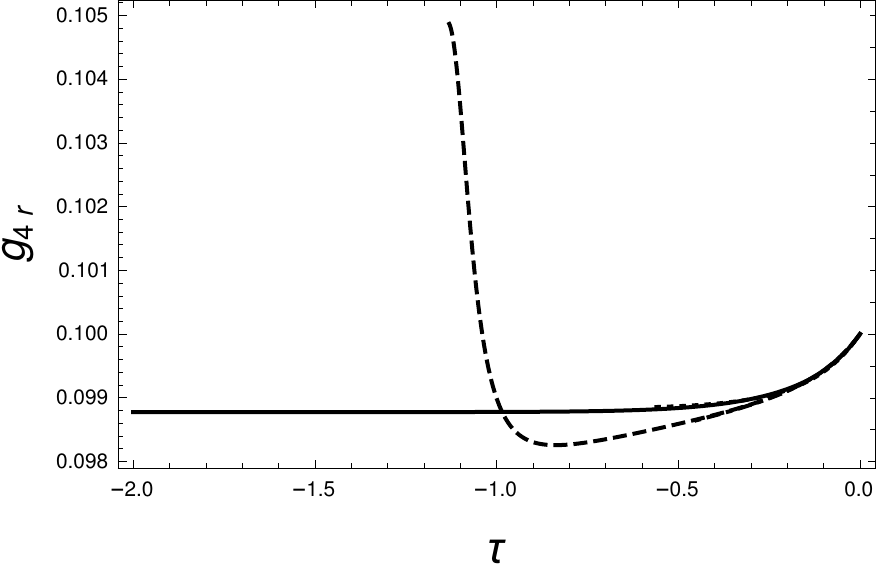}\hskip.5cm\includegraphics[scale=.7]{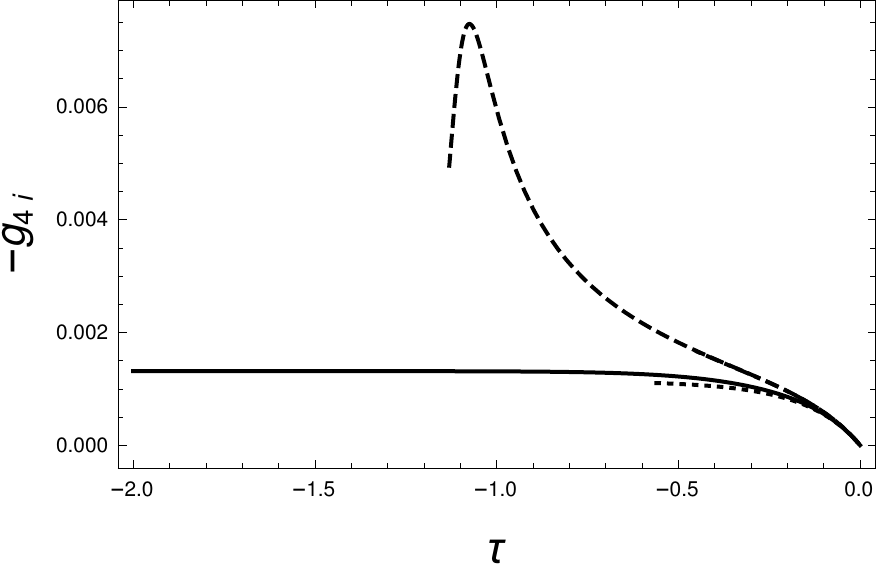}

\includegraphics[scale=.7]{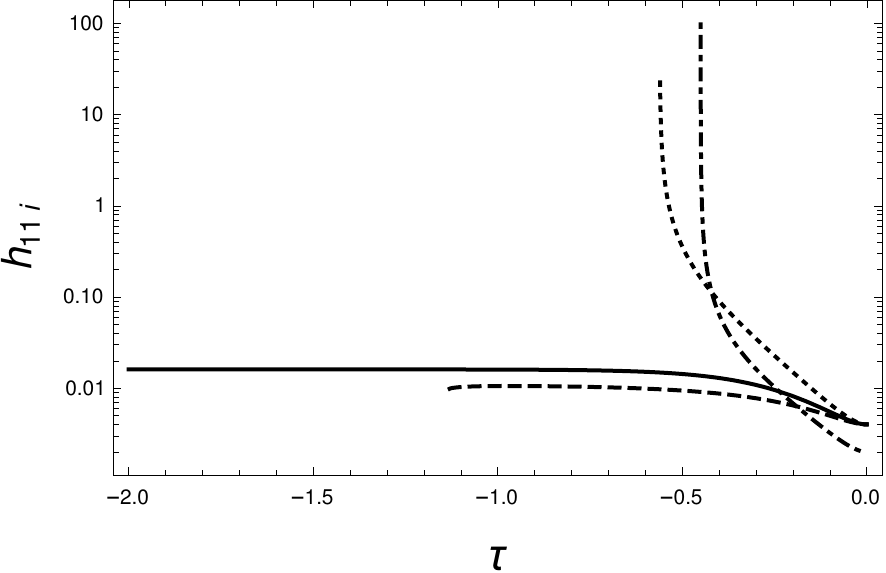}\hskip.5cm\includegraphics[scale=.7]{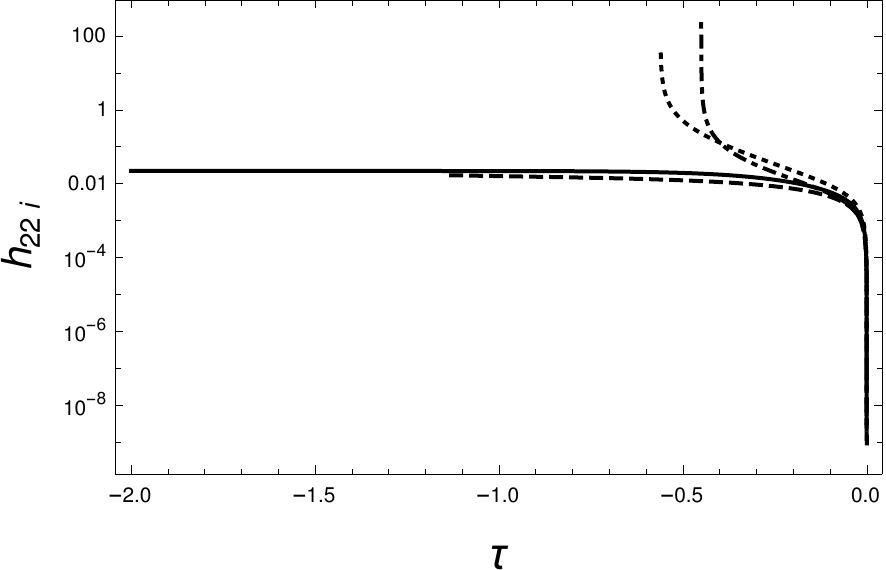}
\caption{Typical trajectories at $g_4(\Lambda)=0.1$, $h_{11}(\Lambda)=-i2g_{2i}(\Lambda)$, $h_{22}(\Lambda)=0$ in the phases $S$ ($g_2(\Lambda)=0.2-i 0.002$, continuous line), $SB$ ($g_2(\Lambda)=-0.006-i 0.002$, dashed line), $S_E$ ($g_2(\Lambda)=0.45-i 0.002$, dotted line) and $SB_E$ ($g_2(\Lambda)=-0.006-i 0.001$, dashed-dotted line), $\eta=1$.}\label{typtrajf}
\end{figure}

\begin{figure}
\includegraphics[scale=.7]{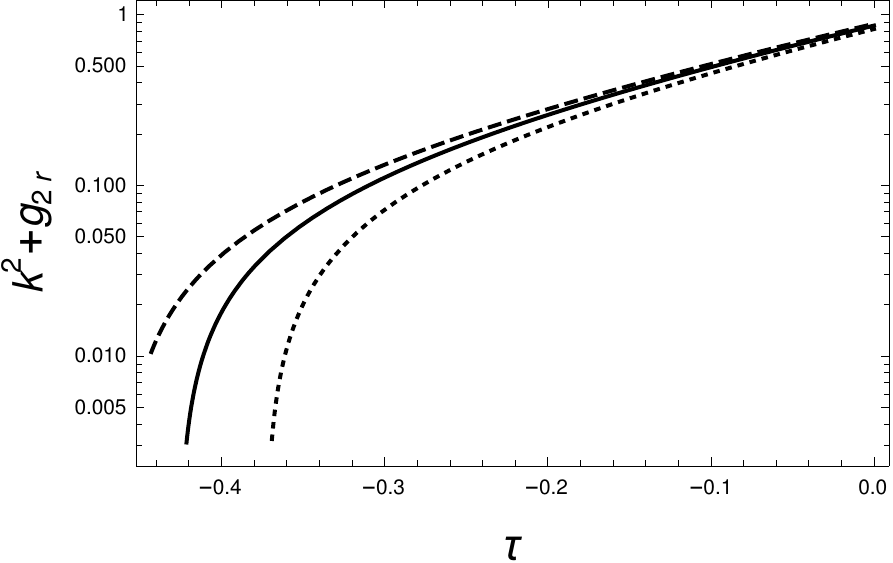}\hskip.5cm\includegraphics[scale=.7]{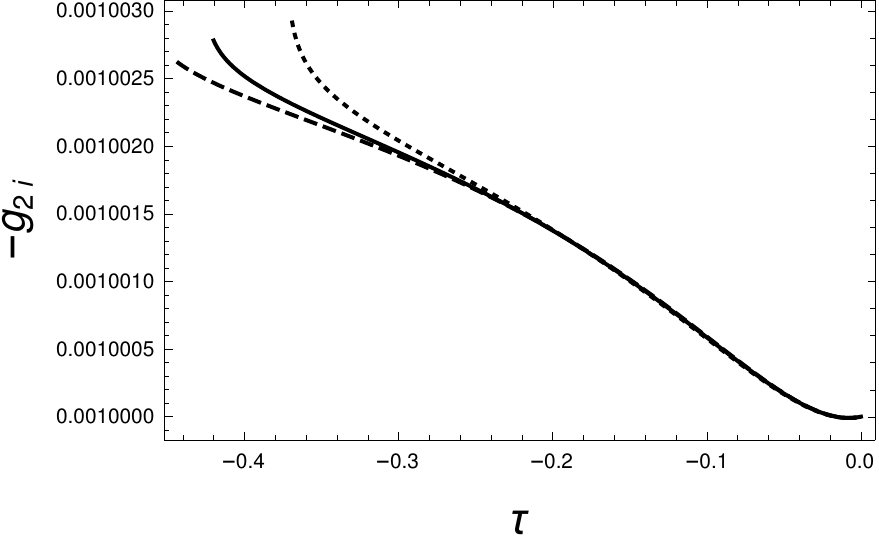}

\includegraphics[scale=.7]{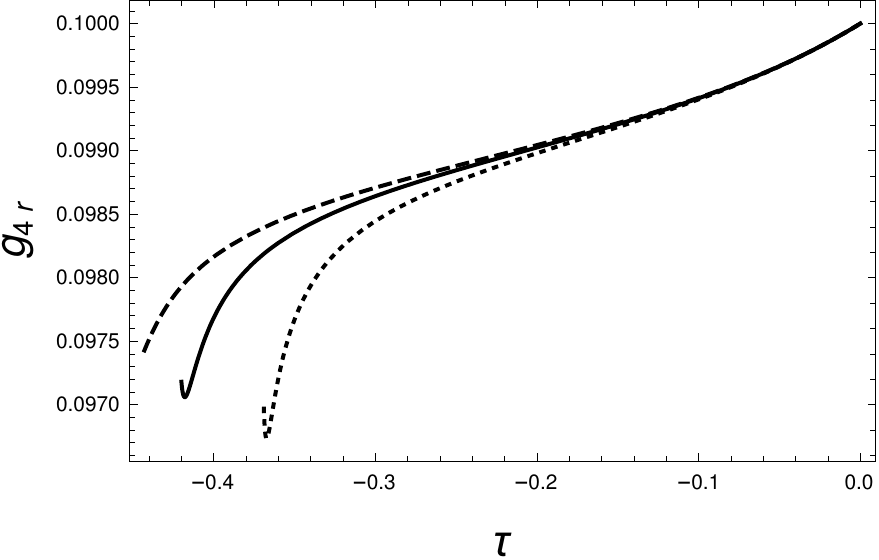}\hskip.5cm\includegraphics[scale=.7]{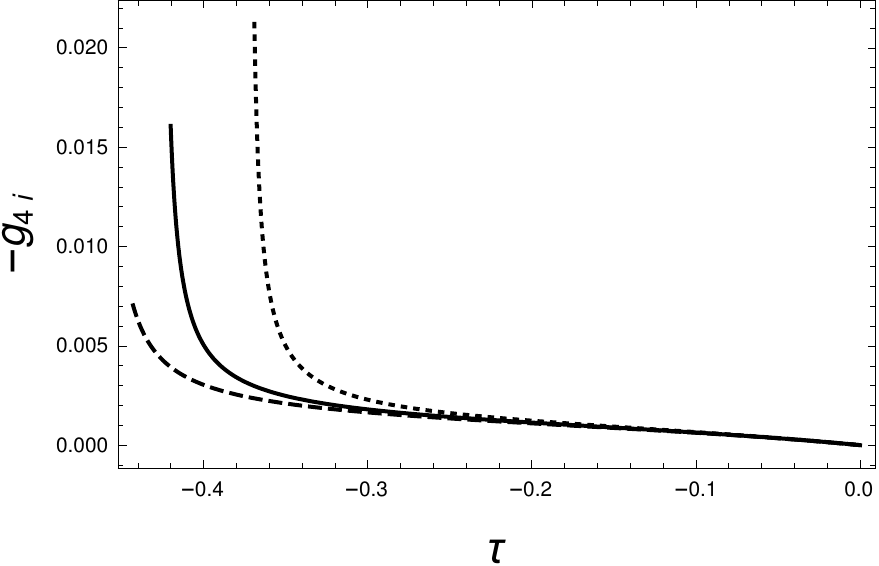}

\includegraphics[scale=.7]{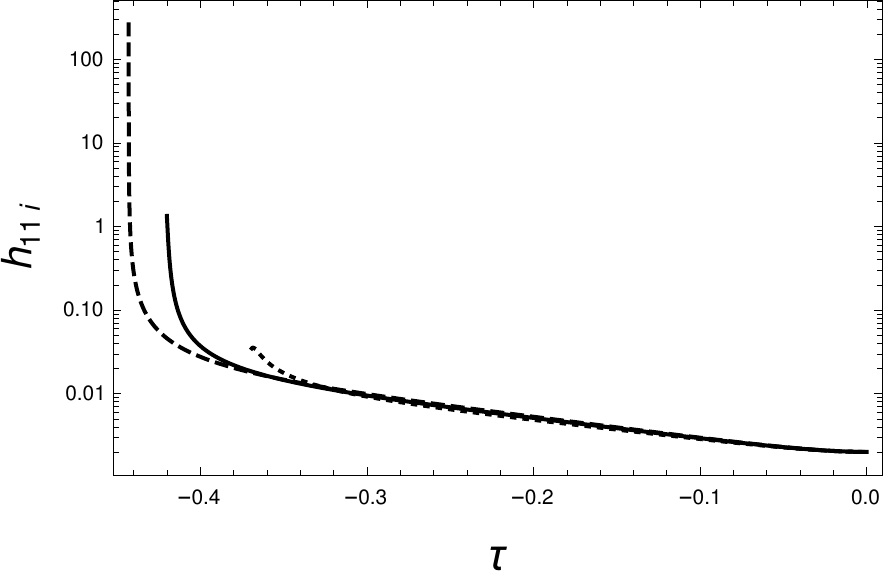}\hskip.5cm\includegraphics[scale=.7]{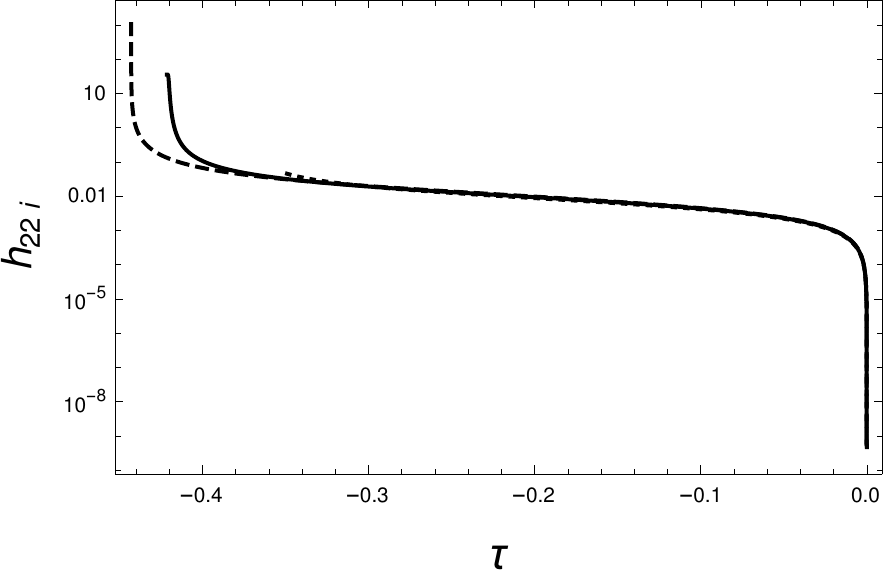}
\caption{Zooming into the $SB-SB_E$ phase boundary. The~trajectories in the $SB$ (dotted line) and $SB_E$ (dashed line) phases together with the separatrix (continuous line) belong to the initial conditions $g_{2r}(\Lambda)=-0.12$ (dashed line), $g_{2r}(\Lambda)=-0.141$ (solid line), $g_{2r}(\Lambda)=-0.18$ (dotted line), $g_{2i}(\Lambda)=-0.001$, $g_4(\Lambda)=0.1$, $h_{11}(\Lambda)=-i2g_{2i}(\Lambda)$, $h_{22}(\Lambda)=0$ and $\eta=1$.}\label{sepf}
\end{figure}

The typical trajectories, shown in Figures~\ref{typtrajf}, indicate the presence of two independent phase transitions, one for the closed and the other for the open parameters. The~left inequality of \eq{intlphs} is satisfied in the entangled phase indicating the presence of strong decoherence. A~more detailed flow at the $SB-SB_E$ phase boundary is given in Figure~\ref{sepf}. The~strong increase of $-g_{2i}$ and $-g_{4i}$ in the $SB$ phase before the evolution has to be halted indicates that the quasi particle become unstable at the onset of the spinodal instability. On~the other side of the phase transition, in~the $SB_E$ phase, the~singularity appears only in the open~parameters. 

\unskip

\unskip

\subsection{Closed Initial~Dynamics}\label{closedts}
The main message of this work is the necessity of using open quantum field theories and retaining the inevitable UV-IR entanglement during the renormalization; see point (i) in the Introduction. While it is well known that the system-environment entanglement plays a decisive role in open quantum dynamics this feature has not been followed in quantum field theory. One can already see from the qualitative picture of Section~\ref{uvirscs} that the movement with the cutoff either towards the IR or the UV direction leads to the accumulation of the entanglement contributions. A~more detailed view of the generation of the mixed contributions to the blocked action can be found by inspecting the renormalization group flow in the limit of a closed initial theory, $\epsilon=-g_{2i}=h_{11i}/2\to0$ and $g_{4i}=h_{22i}=0$.

Let us make a blocking step $k\to k-\dk$ in a closed theory with infinitesimal $\eta$ when $\phi^{(s)}(x)=(e^{i\omega_st}+e^{-i\omega_st})/2$ and consider the beta function of $g_4$. The~energy flowing through the third Feynman graph of Figure~\ref{stpbff} is $\bar\omega=\pm2\omega_s$ or $\bar\omega=0$. The~integrand of the loop integral with integral variable $\omega$ contains two propagators with energies $\omega$ and $\bar\omega-\omega$ in the second line of Equation \eq{loopint}. The~imaginary Dirac-delta peaks of the propagators, written by the help of the identity $1/(x+i\epsilon)=P1/x-i\pi\delta(x)$, coincide for $c_s=1$, $\bar\omega=\pm2\omega_s$ at $\omega=\pm\omega_s$ and generate an $\ord{\epsilon^{-1}}$ imaginary contribution to the beta function. The~same holds for $1\le c_s<1+\ord{\epsilon}$ however the imaginary part drops to $\ord{\epsilon}$ when $1+\ord{\epsilon^0}<c_s$. Such a threshold singularity renders the subtraction point dominated by the propagating particles, $1\le c_s<1+\ord{\epsilon}$, difficult to use~numerically. 

One can avoid the threshold singularity by the use of a subtraction point with finite $\eta$ which is a wave packet rather that a monochromatic wave. The~coinciding poles of the closed theory still generate $\ord{\epsilon^{-1}}$ imaginary contribution to the beta functions of $g_4$ and $h_{22}$ but the dependence on the subtraction point, on~the $c_s$ parameter, is now regular. We could follow the trajectories down to $\epsilon\sim10^{-7}$ with $c_s=1$ and $\eta$ being at least around 10 however the roundoff errors in the initial phase of the evolution arising from the incomplete cancellation between the $\ord{\epsilon^{-1}}$ partial fractions of the loop integral make the trajectory unreliable beyond this~limit. 

\begin{figure}
\includegraphics[scale=.7]{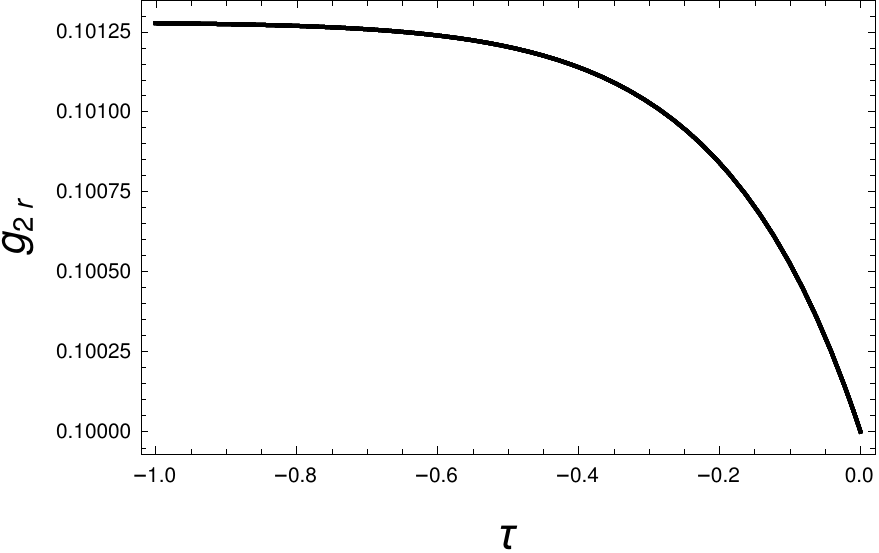}\hskip.5cm\includegraphics[scale=.7]{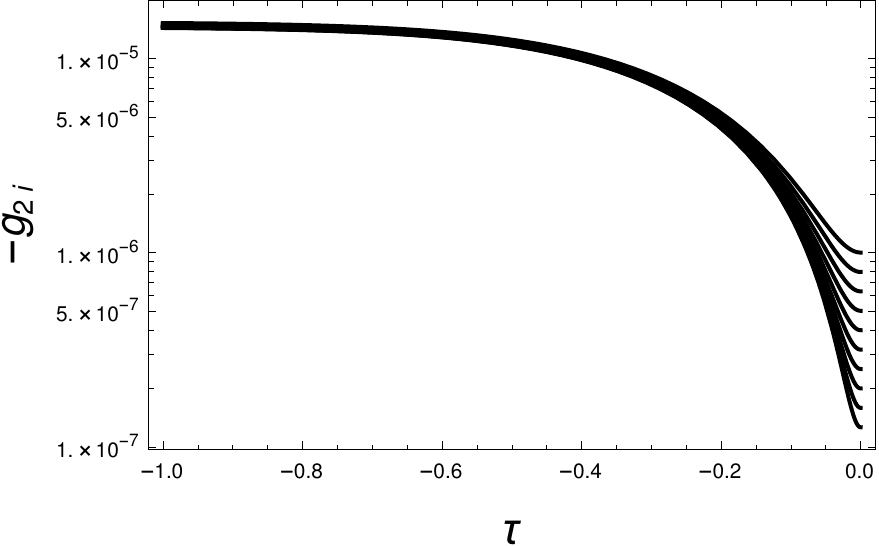}

\includegraphics[scale=.7]{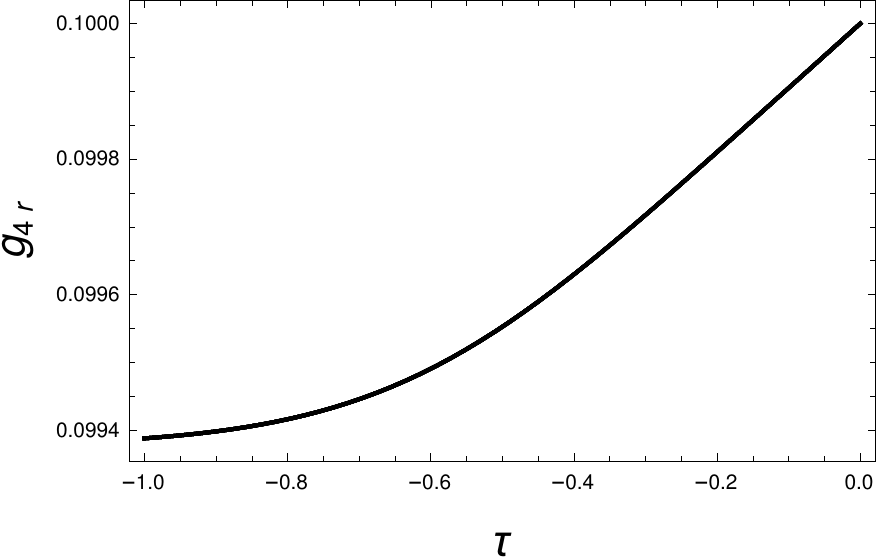}\hskip.5cm\includegraphics[scale=.7]{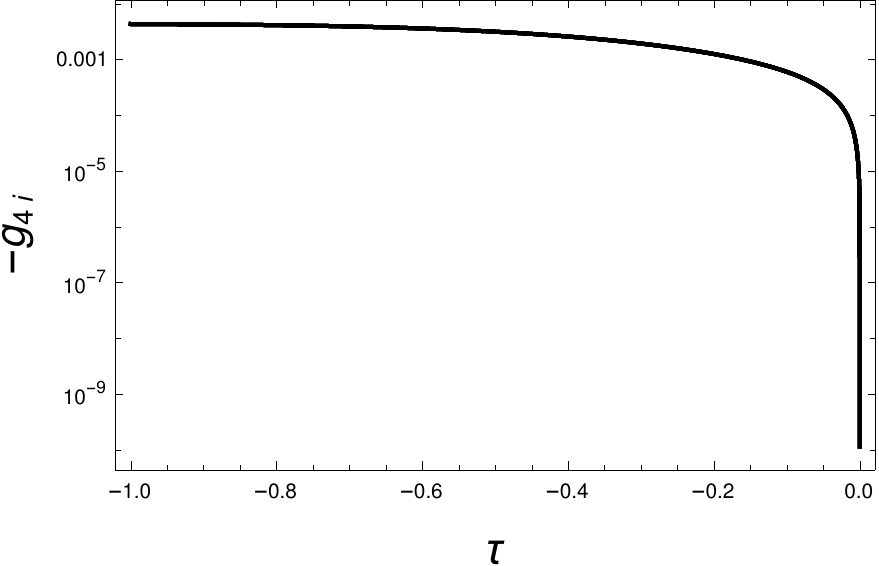}

\includegraphics[scale=.7]{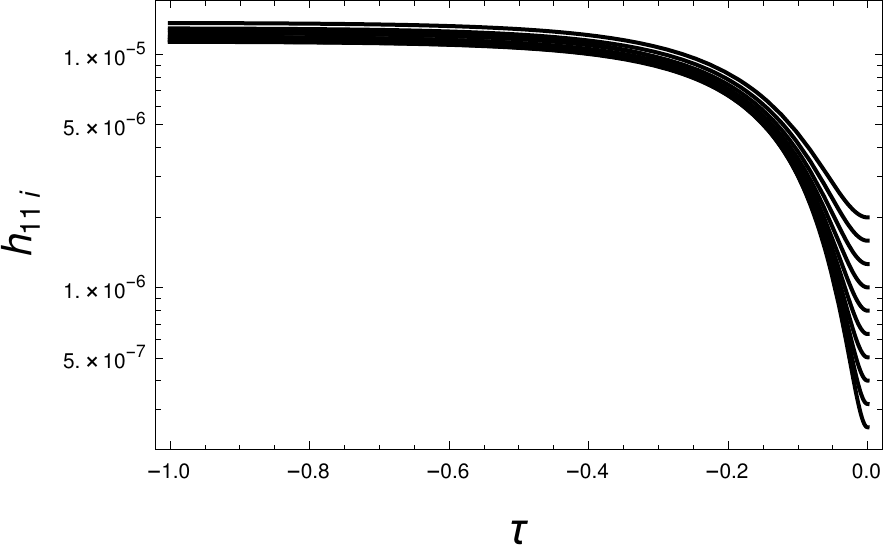}\hskip.5cm\includegraphics[scale=.7]{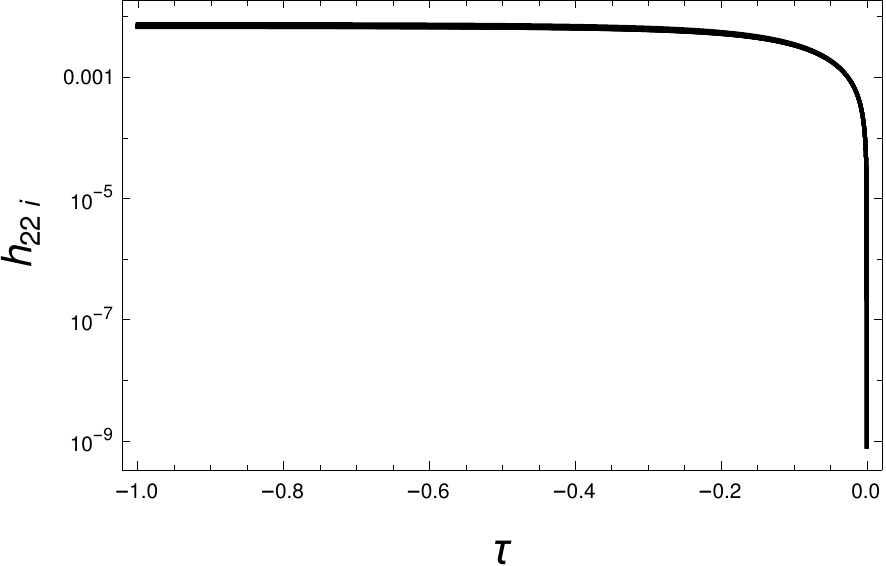}
\caption{The closed limit, $\epsilon\to0$, followed in the interval $-10^{-6}<\epsilon<-10^{-7}$ in the symmetric phase at $g_2(\Lambda)=0.1-i \epsilon$, $g_4(\Lambda)=0.1$, $h_{11}(\Lambda)=-i 2 \epsilon$, $h_{22}(\Lambda)=0$, $\eta=9.2$.}\label{closedf}
\end{figure}

The renormalized trajectory of the symmetric non-entengled phase is shown as $\epsilon$ and is decreased in Figure~\ref{closedf}. The~limit is best tested by the convergence of $g_{2r}$ or $h_{11i}$. The~almost vertical evolution of $g_{4i}$ and $h_{22i}$, the~parameters with vanishing initial condition, is an artifact of the logarithmic plot. In~terms of elementary processes, the second graph of Figure~\ref{ctpbff} drives a rapid increase of $h_{22i}$ by starting in a closed theory which feeds back to accelerate the increase of the originally infinitesimal $h_{11i}$. These two processes are represented by the coefficient $C,D$ of the exponential function in the potential \eq{pote}.

\unskip

\subsection{Relevance of Open~Channels}\label{relopc}
The second quantitative point showing the importance of the open interaction channels consists of an estimate of their impact on the expectation values of physical quantities. Here we face the issue of the relevance of the IR-UV entanglement for observables defined at a scale far below the cutoff scale where the observed system and its environment are separated. A~simple power counting argument indicates that the parameters $h_{11}$ and $h_{22}$ are renormalizable and therefore should be kept in the~action. 

\begin{figure}
\includegraphics[scale=1]{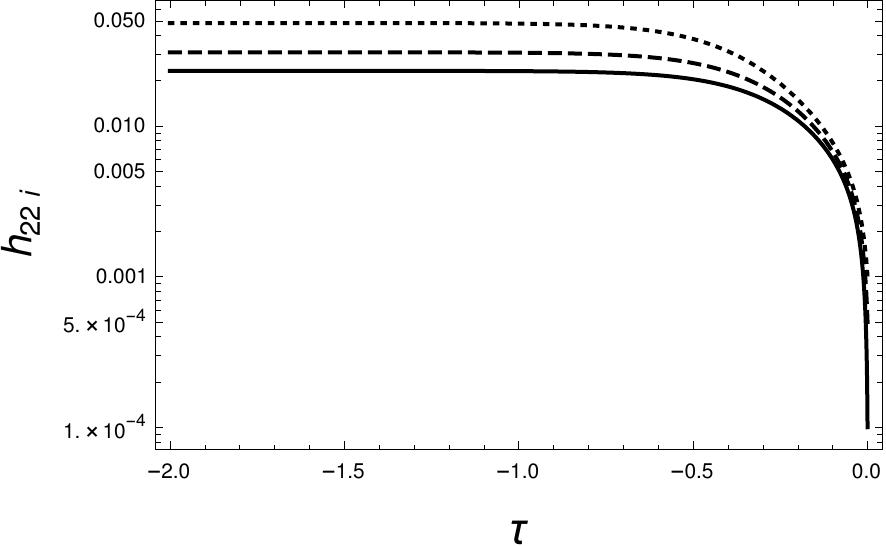}
\caption{Scaling of the open parameters at $g_{2}(\Lambda)=0.2-i 0.002$, $g_{4}(\Lambda)=0.1$, $h_{11}(\Lambda)=-i 2 g_{2i}$, $h_{22}(\Lambda)=i 0.0001$ (solid line), $h_{22}(\Lambda)=i 0.0005$ (dashed line), $h_{22}(\Lambda)=i 0.001$ (dotted line), $\eta=1$.}\label{relf}
\end{figure}

A more detailed view of the mixed contributions to the blocked action can be found by inspecting the renormalized trajectory. The~open channels bring in new two parameters into the action, $h_{11}$ and $h_{22}$. The~former is relevant (super renormalizable) according to power counting. The~latter is marginal and higher order contributions make it relevant or irrelevant (non-renormalizable). To~see what happens, we followed the evolution of $h_{22i}$ corresponding to initial conditions where only its initial condition was slightly changed around zero. The~trajectories plotted on Figure~\ref{relf} correspond to the initial conditions $h_{22i}(\Lambda)=i0.0001,~i0.0005,~i0.001$ and their increasing separation in the UV scaling regime indicates the relevance of this coupling constant, as in point (ii) of the~Introduction.

\unskip

\subsection{Towards the~UV}
Finally, a few words about the UV direction and the issue of renormalizability. The~conditions of  renormalizability can be imposed on four, increasingly restrictive, levels. (i) The cutoff can be sent to infinity without encountering divergences. The~perturbative condition is given by power counting and the $\phi^4$ model of our ansatz belongs to this class. (ii) The renormalization conditions, a~set of non-linear equations, are soluble for the bare parameters. These conditions are violated in non-asymptotically free theories which are restricted to free theories as the cutoff is removed. The~simple qualitative view of the renormalization group flow of the open parameters, mentioned in Section~\ref{uvirscs}, suggests an UV Landau pole for $h_{22i}$ since the friction term with its ``wrong'', unstable sign sends $h_{22i}$ to infinity at finite scale. (iii) The last condition can be strengthened by requiring the stability of the dynamics. The~stability is expressed in terms of inequalities for the imaginary parts of some running parameters, which are specially difficult to maintain. All the trajectories encountered in our numerical efforts to follow the theory in the UV direction led to a condensate, $g_{2i}<0$, or~to the violation of the stability conditions. (iv) The cutoff is assumed to be very large in the multiplicative renormalization group scheme where the contributions, proportional to a negative powers of the cutoff, are neglected. This approximation is untenable in effective theories { where the possibility of placing freely the cutoff to higher energy is assured by requiring the absence of the UV Landau pole. Since the functional renormalization group method handles all contribution within the restricted functional space it is better suited to test the renormalizability}.

One can see without going into the details that the renormalization procedure and the interpretation of possible UV fixed points of open theories remains a challenging open question at the present time.

\section{Summary}\label{concl}
The functional renormalization group equation of the four-dimensional $\phi^4$ model is discussed here to clarify the importance of the open channels of a quantum field theory. The~solution of the evolution equation is sought within a rather simple functional space for the action where the local vertices in space are kept up to fourth order in the field with the minimal necessary time dependence for the open channels. Furthermore, the~absence of condensate is assumed in deriving the evolution~equation.

It is argued that open interactions arise when the cutoff is moved either in the UV or the IR direction. Furthermore, it is found that the open parameters of the action are relevant around the Gaussian fixed point. Thus, closed theories are simply excluded from considerations by requiring an adjustable separation between the observed IR and the unresolved UV degrees of~freedom.

Another result which is important in establishing a relation between the macroscpic and the microscopic physics is that our simple model exhibits a non-perturbative relation between the microscopic (UV) and the strongly decohered macroscopic (IR) degrees of freedom within a closed or almost closed system. Thus, one cannot take the correspondence principle between the classical and the quantum degrees of freedom for~granted.

These results obviously raise further questions. An~obvious issue is the systematic extension of the anstaz space for the action and the check of the stability of these results. This direction requires the use of multi-local actions~\cite{biloc} and the increase of the order of the truncation in the field amplitude. Another question is the boost invariance. The~separation of the degrees of freedom into IR and UV classes is based on the de Broglie wavelength and is not boost invariant. Furthermore, there is a conflict between regulators and boost symmetry since the latter has infinite volume~\cite{boost}. Hence the question arises whether Lorentz symmetry can be maintained in a quantum field theory at any scale. Furthermore, the obvious importance of the system-environment entanglement in open dynamics results in the spectacular success of quantum field theory, by ignoring the open interaction channels rather surprising. A~revisiting of the renormalization program of realistic models is needed to locate the mechanism which operates in certain observations and suppresses the UV-IR~entanglement.

\end{document}